\documentclass[12pt,a4paper]{article} \usepackage[latin1] {inputenc}
\usepackage{amsmath} \usepackage{amsfonts} \usepackage{amssymb}
\begin{document}
\hfill HD-THEP-04-08
\begin{center}
\large{\bf Phenomenological parameterization of quintessence}

\normalsize
\vspace{0.5cm}
C. Wetterich\footnote{e-mail: C.Wetterich@thphys.uni-heidelberg.de}\\
\bigskip
Institut  f\"ur Theoretische Physik\\
Universit\"at Heidelberg\\
Philosophenweg 16, D-69120 Heidelberg
\end{center}
\begin{abstract}
We propose a simple phenomenological parameterization of quint\-essence with a time-varying
equation of state. In particular,
it accounts for the possibility of early dark energy. The quintessence potential can be
reconstructed in terms of the present fraction in dark energy, the present equation
of state and the amount of early dark energy.
\end{abstract}

\bigskip
Distinguishing quintessence - a time varying dark energy component - from a cosmological
constant is a major quest of present observational cosmology. The first
particle physics models \cite{CWQ,RP} are based on a scalar field
rolling down a potential. Typically, a cosmic attractor solution \cite{CWQ,RP} renders
the evolution of the scalar field independent of the initial conditions ( after a certain
transition time in early cosmology). For a given scalar potential the quintessence
cosmology is as predictive as a cosmological constant scenario. However,
there are numerous models with different
potentials. For cosmological tests of their general features one would like a simple
parameterization of the cosmological dy\-na\-mics, say the Hubble parameter as a function
of redshift, $H(z)$, that can later be translated into statements about the form of the
quintessence potential.

A first obviously important parameter is the present fraction in homogeneous dark energy
\cite{CWQ} $\Omega^{(0)}_h=\Omega_h(z=0)$. For observations at low $z$ the next important
parameter is the first derivative of $\Omega_h(z)$ or the present equation of state
$w_0=w_h(z=0)$ \cite{CS}. These two quantities are directly related, cf. eq. (\ref{3}) below.
Continuing an expansion in $z$
is not very meaningful, however, if one wants to cover the physics at high $z$,
like last scattering. For example, the CMB-anisotropies are very sensitive to the fraction
of dark energy at last scattering \cite{Do} $\Omega^{(ls)}_h=\Omega_h(z=1100)$ and
structure formation depends crucially on the weighted mean of $\Omega_h$ during
structure formation $\bar{\Omega}^{(sf)}_h$ \cite{FJ,Sch}. These two important parameters
are not reasonably described by a Taylor expansion of $\Omega_h(z)$ or $w_h(z)$ around $z=0$.

The long term goal is certainly to gather as much information as possible about the whole
function $\Omega_h(z)$ \cite{DW}. One possible strategy leaves $\Omega_h(z)$ essentially
unconstrained and simply tries to find a function which improves substantially the fit to
the available data as compared to a cosmological constant. A complementary
strategy describes $\Omega_h(z)$ \cite{Do, Sch} or $w_h(z)$
\cite{Li, Es, X, CW1} in terms of a few parameters. In particular, the parameterization
of $w_h$ by Fermi-Dirac functions \cite{X} can cover many interesting proposed
quintesssence models. We suggest here a concentration on $\Omega_h$ and on the
absolute minimum of parameters: we propose to
test a three parameter family of quintessence models characterized
(for a flat universe) by $\Omega_M=1-\Omega^{(0)}_h,w_0$ and a new bending parameter $b$
or, equivalently, the fraction of early dark energy, $\Omega_e$. The criteria for the
parameterization are chosen such that at least for a certain parameter range the model is
consistent with a scalar field rolling down a potential. (This is not the case, for example,
for an ansatz $w(z)=w_0+w'z$ except for $w'=0$.)
It also should cover the interesting case of early
quintessence. The two parameters $1+w_0$ and $b$ measure the deviation from the case
of a cosmological constant. At least in principle they can be determined from every
cosmological observation separately since our parameterization covers the whole available
redshift range. If, at the end, values of $1+w_0$ and $b$ different from zero are preferred
and their best values differ between observations covering different
redshift ranges, it will still be time to enlarge our parameterization.

\bigskip
\noindent
{\bf Three parameters}

\medskip
\normalsize Observations have most direct access to the
redshift dependence of the Hubble parameter,
$H(z)$. The dark energy quantity that is most directly related to $H(z)$ is the fraction
in dark energy as a function of redshift, $\Omega_h(z)$. We aim here at a useful
parameterization of $\Omega_h(z)$ in terms of three (or four) parameters. Two parameters
are chosen as $\Omega_M=1-\Omega^{(0)}_h$ and $w_0$.
In order to proceed further we use a simple relation between $\Omega_h(z)$ and
$w_h(z)$ which is valid if the energy density not contained in dark energy can be accounted
for by pressureless matter \cite{CW1}
\begin{equation}\label{1}
\frac{d\Omega_h}{dy}=3\Omega_h(1-\Omega_h)w_h \qquad,\qquad y=\ln(1+z)=-\ln a.
\end{equation}
This suggests to parameterize the function
\begin{equation}\label{2}
R(y)=\ln\left(\frac{\Omega_h(y)}{1-\Omega_h(y)}\right)
\end{equation}
which obeys (in absence of radiation)
\begin{equation}\label{3}
\frac{\partial R(y)}{\partial y}=3w_h(y).
\end{equation}

We propose
\begin{equation}\label{4}
R(y)=R_0+\frac{3w_0y}{1+by}
\end{equation}
where $R_0$ is directly related to $\Omega_M$ by
\begin{equation}\label{5}
R_0=\ln\left(\frac{1-\Omega_M}{\Omega_M}\right).
\end{equation}
The new ``bending parameter'' $b=1/y_b=1/\ln(1+z_b)$ characterizes the redshift where an
approximately constant equation of state turns over to a different behavior. Nonzero $b$
signals the breakdown of the linear expansion for $R(y)$. Within the simple parameterization
(\ref{4}) a positive $b$ is directly related to the presence of early dark energy
\begin{equation}\label{6}
\Omega_e=\Omega_h(y\rightarrow\infty)=\frac{\exp(R_0+3w_0/b)}{1+\exp(R_0+3w_0/b)}.
\end{equation}
Whereas for supernovae (and other observations at low $z$) $b$ is perhaps the most
natural parameter we actually suggest to quote the triplet $(\Omega_M,w_0,\Omega_e)$.
The parameter $\Omega_e$ has a simple physics interpretation. For practical
purposes within the three parameter parameterization $\Omega_e$ equals $\Omega^{(ls)}_h$
and $\bar{\Omega}^{(sf)}_h$ and enters therefore very directly the CMB anisotropies
or structure formation. Inserting eq. (\ref{6}) $b$ is then determined as
\begin{equation}\label{6XA}
b=-\frac{3w_0}{\ln\left(\frac{1-\Omega_e}{\Omega_e}\right)+\ln
\left(\frac{1-\Omega_M}{\Omega_M}\right)}.
\end{equation}
(Nevertheless, for the fitting to the data $b$ is preferred since eq. (\ref{4}) also
covers the case of negative $b$.
Supernovae observe moderate values of $y \simeq 1$ and the parameter $b$ may be related to
$\Omega_h(z=1)$ which is perhaps most directly probed.) For a cosmological constant
one has $w_0=-1,\Omega_e=0$ or $w_0=-1,~b=0$
whereas a constant equation of state obtains for $b=0$. If one
performs a Taylor expansion of $w_h(z)$ \cite{Li}
\begin{equation}\label{7}
w_h(z)=w_0+w'z+\dots
\end{equation}
one finds the relation
\begin{equation}\label{8}
w'=-2w_0b.
\end{equation}

Neglecting radiation, the function $R(y)$ has a simple interpretation in terms of the
averaged equation of state
\begin{equation}\label{8A}
\bar{w}_h(y)=\frac{1}{y}\int\limits^y_0 dy'w_h(y')=
\frac{R(y)-R_0}{3y}.
\end{equation}
This yields a very convenient formula for the redshift dependence of the Hubble
parameter according to
\begin{equation}\label{8B}
\frac{H^2(z)}{H^2_0}=(1-\Omega_M)
(1+z)^{3+3\bar{w}_h(z)}
+\Omega_M(1+z)^3.
\end{equation}
Therefore the luminosity distance and related cosmological quantities are easily
expressed in terms of our parameterization of $R$. With (\ref{4}) one finds
\begin{equation}\label{8C}
\bar{w}_h(z)=\frac{w_0}{1+b\ln(1+z)}.
\end{equation}
Our ansatz can also be interpreted as a direct parameterization of the redshift dependence
of the Hubble parameter.

We note that our parameterization ensures automatically that $\Omega_h(z)$ remains between
$0$ and $1$ for arbitrary values of $R(y)$. It implies a negative equation of state (for
$w_0<0)$
\begin{equation}\label{9}
w_h(y)=\frac{w_0}{(1+by)^2}.
\end{equation}
Therefore $R(y)$ and $\Omega_h(z)$ are monotonic functions. For quintessence due to the
evolution of a scalar field one has
\begin{equation}\label{10}
w_h=\frac{T-V}{T+V}
\end{equation}
with kinetic and potential energy $T$ and $V$. For positive $T$ and $V$ the equation of
state is bounded to the interval $-1\leq w_h\leq 1.$ (For $T>0,~V<0$ one finds $|w_h|>1$,
with sign depending on the sign of $T+V$.)
One may argue that the form (\ref{9}) is not general enough and, in particular, not
suitable for the description of a possible change of sign of $w_h$. In fact, for
typical models with early dark energy one expects $w_h>0$ in the radiation dominated
epoch. We remark, however, that eqs. (\ref{1}),(\ref{3}) receive corrections \cite{CW1}
once radiation becomes important. (For $\Omega_e>0$ one finds for the radiation
dominated epoch $w_h=1/3$.) Our parameterization can therefore remain reasonable as
long as $\Omega_h(z)$ is a monotonic function
\footnote{Actually, in many models the potential is essentially exponential at early times.
In this case a calculable change \cite{CWQ} in $\Omega_h(z)$ occurs at the transition
from the radiation to the matter dominated era. Beyond $\Omega_e$ this jump does not involve
a new free parameter and could easily be incorporated into our parameterization.
Since we concentrate here on the matter dominated epoch we have left out this issue in this
note.} which is the case for most quintessence models.

\bigskip
\noindent
{\bf Reconstructing the scalar potential}

\medskip
The scalar potential and kinetic term can be reconstructed \cite{CW1} from $R(y)$.
As a consequence, our three parameters describe a family of complete models for which
not only the background evolution but also the fluctuations of the cosmon field can be
determined. One first constructs $V(y)$ according to
\begin{equation}\label{12}
V=\frac{1-w_h}{2}\rho_h=\frac{3\bar{M}^2}{2}
(1-w_h)\Omega_hH^2
\end{equation}
(with reduced Planck mass $\bar{M}^2=M^2_p/8\pi,~\rho_{cr}=3\bar{M}^2H^2)$. For the relation
between the scalar field $\varphi$ and $y$ we also use the scalar kinetic energy
\begin{eqnarray}\label{13}
T&=&\frac{3\bar{M}^2}{2}(1+w_h)\Omega_hH^2\nonumber\\
&=&\frac{1}{2}k^2(\varphi)\dot{\varphi}^2=\frac{k^2}{2}
\left(\frac{\partial\varphi}{\partial y}\right)^2\dot{y}^2=
\frac{k^2}{2}H^2\left(\frac{\partial\varphi}{\partial y}\right)^2.
\end{eqnarray}
One possibility employs a standard kinetic term $(k=1)$ and integrates the relation
\begin{equation}\label{14}
\frac{1}{\bar{M}}\frac{\partial\varphi}{\partial y}=\sqrt{3(1+w_h)\Omega_h}
\end{equation}
in order to obtain $\varphi(y)$, and, with (\ref{12}{), $V(\varphi)$.
Perhaps more direct and more elegant is a rescaling $\varphi=\varphi(\phi)$ such that the
potential $V(\phi)$ takes a standard form. The information about the specific quintessence
model is then contained in the ``kinetial'' $k(\phi)$ \cite{HW}.
In this case $\phi(y)$ can be
directly extracted from eq. (\ref{12}). Subsequently, eq. (\ref{13}) can be solved for $k$.
In particular, if the potential is monotonic in the scalar field one may use the
``standard exponential form''
\begin{equation}\label{15}
V=\bar{M}^4\exp\left(-\frac{\phi}{\bar{M}}\right)
\end{equation}
For this choice one extracts from eqs. (12)(14) of ref. \cite{CW1}
\begin{equation}\label{16}
k^{-1}=\left [3(1+w_h)\Omega_h\right]^{-1/2}\left\{3(1+w_h)-\frac{\partial w_h}{\partial y}
\frac{1}{1-w_h}\right\}
\end{equation}
and from eqs. (\ref{12}) (\ref{15})
\begin{equation}\label{20XA}
\frac{\phi}{\bar{M}}=-\ln
\left(\frac{3H^2}{2\bar{M}^2}(1-w_h)\Omega_h\right).
\end{equation}
In conclusion, we can associate to a given parameter set $(\Omega_M,w_0,b)$ a
quintessence model with a specific kinetical $k(\phi)$, or after rescaling to a standard
kinetic term, a potential $V(\varphi)$.

A few comments illustrate this relation.
\begin{itemize}
\item [(i)] For $\Omega_h>0$ a positive $T$ requires $w_h\geq -1$.
\item [(ii)] For a region with $\Omega_h>0$ and $w_h>1$ the potential $V$ must be negative
(cf. eq. (\ref{12})). The parameterization (\ref{4}) does not cover this case.
\item [(iii)]The relation (\ref{16}) holds only up to a sign. For $k^{-1}=0$ the potential
$V(\varphi)$ has a stationary point, $\partial V/\partial\varphi=0$. A change of sign of
$k^{-1}$ according to eq. (\ref{16}) signals that the scalar field moves through an
extremum of the potential $V(\varphi)$ at the corresponding value of $y$. This implies an
interesting condition for a monotonic $V(\varphi)$, namely \footnote{This condition follows
also directly from eq. (\ref{9}). It is equivalent to $\partial\ln V/\partial y\neq 0$,
using $\partial\ln\rho_h/\partial y=3(1+w_h)$.}
\begin{equation}\label{17}
\frac{\partial w_h}{\partial y}\neq 3(1-w^2_h).
\end{equation}
Within the ansatz (\ref{4})
$k^{-1}$ is positive for large $y$ if $b>0$. A monotonic potential
therefore requires for $y=0$
\begin{equation}\label{18}
\frac{\partial w_h}{\partial y}_{|y=0}<3(1-w^2_0)
\end{equation}
or $(w_0<0)$
\begin{equation}\label{19}
b<-\frac{3(1-w_0^2)}{2w_0}.
\end{equation}
\end{itemize}

Our parameterization covers a wide class of dynamical behaviors of dark energy beyond the
quintessence models based on a scalar field. We recall that $w_0<1$
or $b<0$ are not compatible with scalar quintessence models
(with $T>0$). For $w_0$ close to $-1$, as suggested by observation, only a rather narrow
range of $b$ is compatible with a monotonic potential. As an immediate consequence,
quintessence models with monotonic potential and leading to $w_0$ close to $-1$ are very
hard to distinguish from a cosmological constant by observations at low $z$. Indeed,
$w(z)$ is bounded by the solution of the differential equation $\partial w_h/\partial y=
3(1-w^2_h)$ (cf. eq. (\ref{17})), namely
\begin{equation}\label{20}
w^{up}_h(y)=-\frac{1-w_0-(1+w_0)e^{6y}}{1-w_0+(1+w_0)e^{6y}}.
\end{equation}
Independently of our specific parameterization a monotonic potential requires
$w_h(z)\leq w^{up}_h(z)$. For small $(1+w_0)$ one needs sufficiently large
$z$ before $w(z)$ can deviate substantially from $-1$. Actually, the bound
(\ref{20}) cannot be saturated since this would require that the scalar field is frozen
at an extremum, in contradiction to $w_h>-1$. Combining the bound (\ref{19}) with
eq. (\ref{9}) yields a more severe bound within our parameterization
\begin{equation}\label{21}
w_h(y)<\frac{w_0}{\left(1-\frac{3(1-w^2_0)}{2w_0}y\right)^2}\approx
-\frac{1}{\left[1+3(1+w_0)y\right]^2}
\end{equation}
where the last expression assumes $1+w_0 \ll 1$. In this case, i.e. for $b$ obeying
the inequality (\ref{19}), we find a monotonic $k(\phi)$.

\bigskip
\noindent
{\bf Extended parameterization}

\medskip
Combining the inequality (\ref{19}) with eq. (\ref{6XA}) restricts $\Omega_e$
to rather small values if $1+w_0\ll 1$. This is, of course, a consequence
of our ansatz (\ref{4}) with only three independent parameters. From the particle physics
point of view there is no contradiction between a sizeable $\Omega_e$ (say a few percent)
and $w_0$ close to $-1$. One may therefore want to weaken the strict connection between
$b$ and $\Omega_e$ by an extended parameterization. For example, this may be required if
supernovae find a small $b$ while structure formation or the CMB prefer a nonvanishing
$\Omega_e$. A still relatively simple possibility is the choice
\begin{equation} \label{26}
\Omega_h(y)=\frac{(\tilde{\Omega}_\alpha(y)-\Omega_e)^\alpha}
{(1-\Omega_M-\Omega_e)^{\alpha-1}}+\Omega_e
\end{equation}
where
\begin{equation}
\tilde{\Omega}_\alpha(y)=
\frac{(1-\Omega_M)\exp
\left(\frac{3w_0}{\alpha}\frac{y}{1+by}\right)}
{\Omega_M+(1-\Omega_M)\exp
\left(\frac{3w_0}{\alpha}\frac{y}{1+by}\right)}
\end{equation}
coincides with $\Omega_h(y)$ and the ansatz (\ref{4}) for $\alpha=1$. The relation
between $b$ and $\Omega_e$ depends now on the new parameter $\alpha$
\begin{equation}
b=-\frac{3w_0}{\alpha}
\left[\ln\left(\frac{1-\Omega_e}{\Omega_e}\right)+\ln
\left(\frac{1-\Omega_M}{\Omega_M}\right)\right]^{-1}.
\end{equation}
For a Taylor expansion $w_(z)=w_0+w'z$ one finds
\begin{equation}
w'=-2bw_0+
\frac{3(\alpha-1)}{\alpha}
\frac{\Omega^2_h-\Omega_e(2\Omega_h-1)}{\Omega_h-\Omega_e}w^2_0.
\end{equation}

Another possible shortcoming of the ansatz (\ref{4}) (and also the extension
(\ref{26})) is the
necessarily nonzero value of $\Omega_e$ for $b>0$.
One may include the possibility that $\Omega_h$ does not approach a constant for large
$y$ even for $b\neq 0$ and use an alternative extended parameterization
\begin{equation}\label{11}
R(y)=R_0+\frac{3w_0y+3b\gamma y^2}{1+by}
\end{equation}
with
\begin{equation}\label{11A}
w_h(y)=\frac{w_0+b\gamma y(2+by)}{(1+by)^2}.
\end{equation}
In practice, one may express $\gamma$ in terms of the fraction in dark energy at last
scattering, $\Omega^{(ls)}_h=\Omega_h(z=1100)$.
Negative $\gamma$ implies that $\Omega_h$ approaches
zero for $y\rightarrow\infty$, whereas it formally increases towards one for
$\gamma>0$. Neglecting radiation $\gamma$ denotes the ``asymptotic'' value of
$w_h(z\rightarrow\infty)=\gamma$. (For this particular point we assume
$b>0$ such that $R$ has no pole for
$y>0$.) Extensions of the parameterization like (\ref{26}) or (\ref{11}) may be needed
in the long run for a
detailed comparison of CMB, structure formation and supernovae. For observations covering
a range $y\begin{array}{c}
\vspace{-0.3cm}<\\
\sim
\end{array}1$ the parameterization (\ref{4}) is presumably sufficient.

Other parameterizations of a time varying equation of state $w_h(z)$ or the
corresponding $\Omega_h(z)$ have been proposed in the literature
\cite{Do,Sch,Li,Es,X,CW1}. They often cover only a restricted range of redshift
but fail (or are unspecified) when applied to the full redshift range accessible
to observation, say $0\leq z\begin{array}{c}\vspace{-0.3cm}<\\\sim\end{array}1200$.
For example, the Taylor expansion $w(z)=w_0+w'z$ leads to $w(z)>1$ for
$z>z_{cr}=(1-w_0)/w'~(w'>0)$. This only is consistent with a very restricted and not very
natural class of scalar models, namely those where the potential $V(\varphi)$ was negative
in the past for $z>z_{cr}$ and has turned positive recently for $z<z_{cr}$.
(For negative $w'$ the positivity of kinetic energy $(w(z)>1)$ is violated in the past - even
worse.) On the other hand, the proposed parameterizations covering the whole redshift range
lead usually to a more complicated form of $H(z)$ as compared to
eq. (\ref{8B}), involve more than three parameters
and often contain a certain degree of arbitrariness.

It is obvious that every parameterization in terms of a few parameters will have its
strength for a particular problem but also its shortcomings when applied to the whole
relevant redshift-range. Nevertheless, concentrating on a simple
one would be beneficial for a direct comparison of different
observations at the present stage.

\bigskip
\noindent
{\bf Acknowledgment}

\medskip
The author would like to thank M. Doran, C. M\"uller and G. Sch\"afer for
helpful discussions.

\end{document}